\documentclass[11pt,twoside]{article}
\usepackage{asp2010,graphicx}
\resetcounters
\begin{document}

\title{Infrared Survey of Pulsating Giant Stars in the Spiral Galaxy M\,33:
Dust Production, Star Formation History, and Galactic Structure}
\author{Atefeh~Javadi$^{1,2}$, Jacco~Th.~van~Loon$^2$, and Mohammad Taghi
Mirtorabi$^1$
\affil{$^1$Physics Department, Alzahra University, Vanak, Tehran, Iran}
\affil{$^2$Lennard-Jones Laboratories, Keele University, United Kingdom}}

\begin{abstract}
We introduce a near-IR monitoring campaign of the Local Group spiral 
galaxy M\,33, carried out with the UK IR Telescope (UKIRT). The 
pulsating giant stars are identified and their distributions are used 
to derive the star formation rate as a function of age. We here present 
the star formation history for the central square kiloparsec. These 
stars are also important dust factories; we measure their dust
 production rates from a combination of our data with {\it Spitzer}
Space Telescope mid-IR photometry.
\end{abstract}

\section{Getting to Know the Milky Way from Studies of M\,33}

Given the difficulty of observing the central regions of our own Milky Way
galaxy \citep{vanloon_2003} we might find out more about the structure and
evolution of spiral galaxies by studying other nearby examples. Among the
Local Group spiral galaxies, M\,33 is smaller than the Milky Way and M\,31 
but it is viewed by us more favourably; the distance modulus to M\,33 is
$\mu=24.9$ mag \citep[955 kpc;][]{bonanos_2006}.

\section{Pulsating Giant Stars as Tracers of Star Formation and Dust
Production}

Pulsating giant stars have reached the final stages of their evolution
\citep[their lives cut short by the severe mass loss initiated by these
pulsations;][]{vanloon_2008}. As their luminosity depends on their core 
mass, which depends on the birth mass, pulsating giant stars are good 
tracers of the population of stars formed when they themselves formed. 
Their cool, extended atmospheres are also fertile grounds for the 
formation of dust grains. As the grains intercept visual radiation from 
the star and emit it at infrared (IR) wavelengths we can measure the dust 
production rate by modelling the spectral energy distribution.

\section{The United Kingdom Infrared Telescope Monitoring Survey of M\,33}

\begin{figure}[!ht]
\hspace{10mm}\includegraphics[scale=0.56]{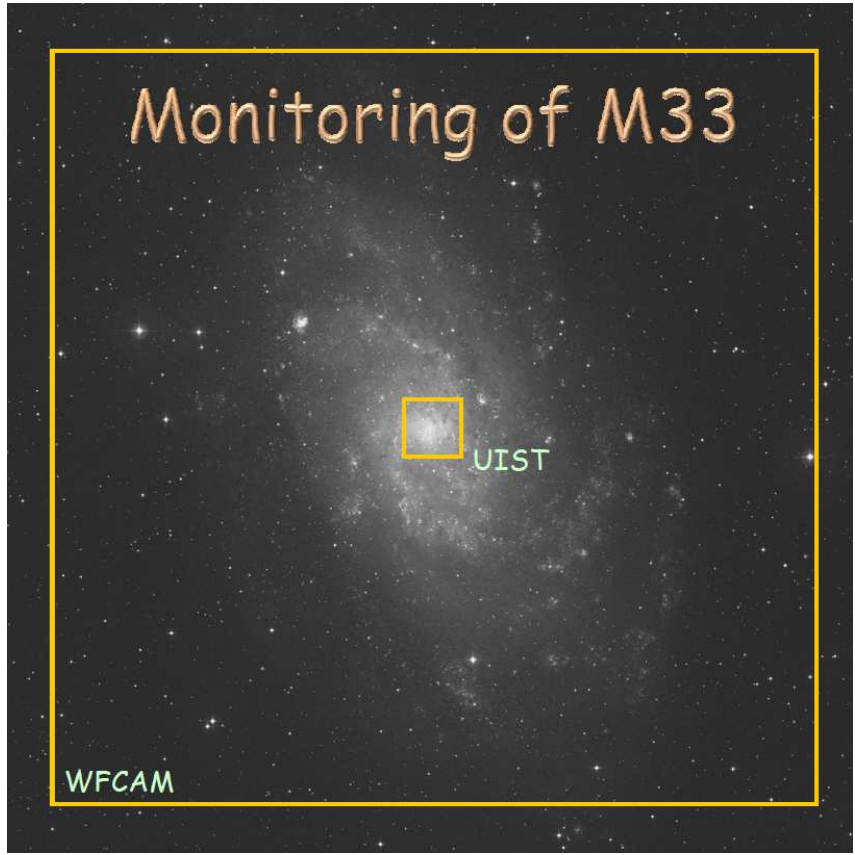}
\caption{Areas covered by our UKIRT monitoring campaigns on the central 
square kiloparsec with UIST, and the M\,33 disc with WFCAM.}
\label{cover}
\end{figure}

We used the United Kingdom Infrared Telescope (UKIRT) to monitor M\,33 
in the $K$ band (at a wavelength of 2.2 $\mu$m). This was done first for 
the central $4^\prime\times4^\prime$ (a square kpc at the distance of 
M\,33), mainly with the UIST instrument over the period from 2003 to 2007. 
Later in the campaign, images were taken with the WFCAM instrument, 
covering essentially the entire extent of the visible disc of M\,33 
(Fig.~\ref{cover}). Occasionally, images were taken also in the $J$ 
(1.2 $\mu$m) and $H$ (1.6 $\mu$m) bands in order to obtain colour 
information. The survey and identification of variable stars are
described in detail in \citet*[Paper I]{javadi_2010}; 812 variable stars 
were found and shown to be predominantly pulsating giant stars -- the full
photometric catalogue comprises 18\,398 stars and is available from CDS.

\section{The Star Formation in the Central Square Kiloparsec of M\,33}

The star formation history is described by the star formation rate, $\xi$, 
as a function of look-back time (``age''), $t$:
\begin{equation}
\xi(t) = \frac{f(K(M(t)))}{\Delta(M(t))f_{\rm IMF}(M(t))},
\end{equation}
where $f(K)$ is the observed $K$-band distribution of pulsating giant stars,
$\Delta$ is the duration of the evolutionary phase during which these stars
display strong radial pulsation, and $f_{\rm IMF}$ is the Initial Mass
Function describing the relative contribution to star formation by stars of
different mass. Each of these functions depends on the stellar mass, $M$, 
and the mass of a pulsating star at the end of its evolution is directly 
related to its age ($t$).

\begin{figure}[!ht]
\hspace{13mm}\includegraphics[scale=0.53]{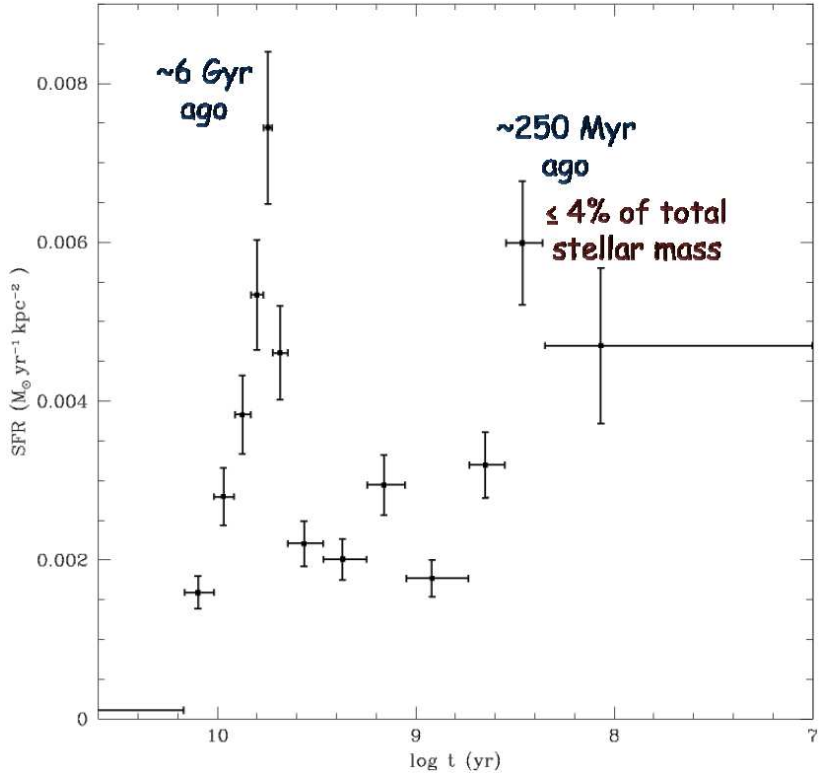}
\caption{The ~star ~formation ~history ~in ~the ~central ~square 
~kiloparsec ~of ~M\,33 derived from pulsating giant stars found in 
our infrared monitoring survey.}
\label{sfh}
\end{figure}

The resulting star formation history is shown in Fig.~\ref{sfh} and is 
described in detail in Paper II (Javadi et al., submitted to MNRAS). 
The main features are that the large majority of the stars were formed 
more than 4 Gyr ago, but that the subsequently quieter star formation 
has been punctuated with epochs of enhanced rates of star formation of 
which a recent one is detected to have occurred around 200--300 Myr ago, 
forming at most 4\% of all stars that have been formed over M\,33's 
lifetime (within the central square kiloparsec).

The spatial distributions of the massive stars, intermediate-age Asymptotic
Giant Branch (AGB) stars and generally old \,Red \,Giant \,Branch \,(RGB) 
\,stars
suggest that young and intermediate-age stars were formed within the disc,
while the oldest stars may inhabit a more dynamically-relaxed configuration.
Interestingly, the massive stars concentrate in an area South of the 
nucleus, and the intermediate-age population shows signs of a 
``pseudo-bulge'' that however may well be a bar-like feature.

\section{Dust Production in the Central Square Kiloparsec of M\,33}

\begin{figure}[!ht]
\hspace{5mm}\includegraphics[scale=0.62]{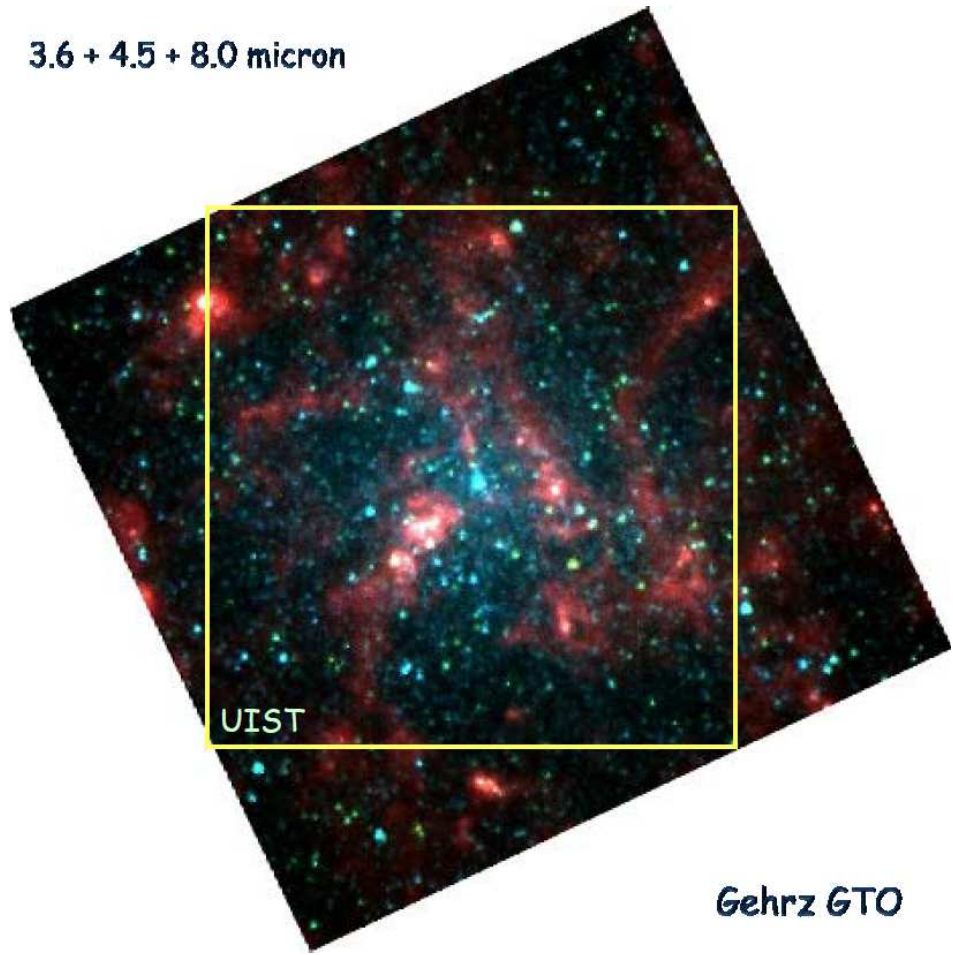}
\caption{The central field overlain on a {\it Spitzer} Space Telescope 
mid-IR composite.}
\label{spitz}
\end{figure}

\begin{figure}[!ht]
\hspace{8mm}\includegraphics[scale=0.61]{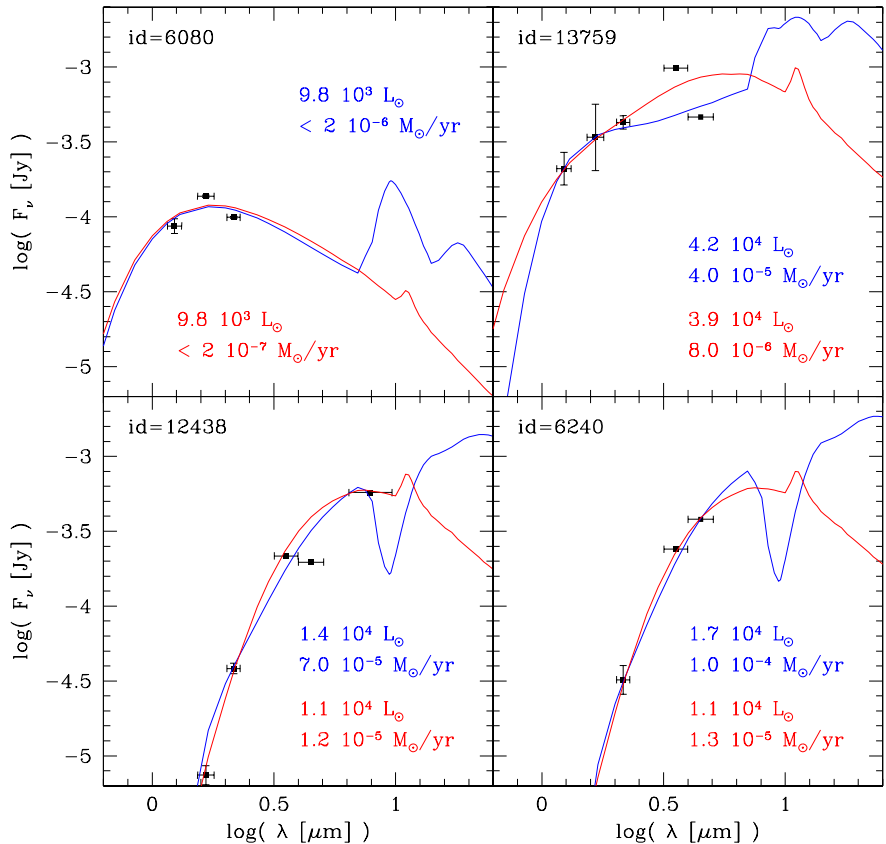}
\caption{UKIRT + {\it Spitzer} (where available) photometry of four 
examples of red giant stars in the centre of M\,33, that are affected 
by various levels of mass loss. The photometry is compared with 
{\sc dusty} radiative transfer models \citep{nenkova_1999} for 
oxygen-rich dust (yielding higher rates and higher luminosities beyond 
10~$\mu$m) and carbon-rich dust. The light curves of these stars are 
presented in Paper I.}
\label{dust}
\vspace*{-3mm}
\end{figure}

Despite the complex diffuse emission and crowdedness in the central 
regions of M\,33, a significant fraction of the pulsating giant stars 
have been detected at mid-IR wavelengths (3--8 $\mu$m) with the 
{\it Spitzer} Space Telescope \citep[Fig.~\ref{spitz}, cf.][]{mcquinn_2007}. 
It has been possible to estimate the mass-loss rates (and dust production 
rates) across a range of intensity (Fig.~\ref{dust}, Paper III in 
preparation).

\section{On-going Work and Concluding Remarks}

We are currently extending our study to the WFCAM data that cover the 
disc of M\,33, to derive a global star formation history and dust 
production rate. We aim to establish a link between the dust return 
and the formation of stars within the prominent spiral arm pattern, 
and to map their subsequent dynamical relaxation into the inter-arm 
regions.

In conclusion, our method to derive the star formation history from 
pulsating giant stars has been validated for the central region of 
M\,33. While model-dependent, our analysis is internally consistent 
and supports the Padova models we employed \citep{marigo_2008} except 
that the super-AGB stars do seem to reach high luminosities and develop 
cool atmospheres and strong pulsation.

\acknowledgements We are grateful for support from the conference 
organisers, and from the excellent and dedicated staff at the United 
Kingdom Infrared Telescope. This project was funded by The Leverhulme 
Trust (grant No.\ RF/4/RFG/2007/0297).

\bibliography{b_vanloon}

\end{document}